\documentclass[12pt]{article}

\usepackage[utf8]{inputenc}
\usepackage{amsmath, amsthm, amssymb, amsfonts}
\usepackage{import}
\usepackage{natbib}
\usepackage{fullpage}
\usepackage{authblk}
\usepackage[bibliography=common]{apxproof}
\usepackage{import}
\usepackage{mathtools}
\usepackage{tikz}
\usepackage{float}
\usepackage{enumitem}
\usepackage{bbm}

\setcitestyle{authoryear,open={(},close={)}}

\usetikzlibrary{calc,patterns,angles,shapes, positioning, intersections, quotes}

\newtheoremrep{thm}{Theorem}
\newtheoremrep{lem}{Lemma}
\newtheoremrep{prop}{Proposition}

\newtheoremrep{proposition}{Proposition}

\usepackage{graphicx}
\newtheorem{corollary}{Corollary}

\newtheorem*{example}{Example}

\theoremstyle{definition}
\newtheorem{definition}{Definition}
\newtheorem{assumption}{Assumption}

\usepackage[colorlinks=false,pagebackref=true, hidelinks]{hyperref}

\newcommand{\R}{\mathbb{R}}

\def\l{\lambda}

\newcommand{\eps}{\varepsilon}
\def\priority{\succcurlyeq}

\def\indic{\mathbbm{1}}

\def\F{\mathcal{F}}
\def\Eq{\mathcal{E}}
\def\Top{\text{Top}_{\priority}}

\def\PO{\mathcal{P}}

\def\E{\mathbb{E}}
\def\P{\mathbb{P}}

\def\given{\, | \,}

\allowdisplaybreaks

\begin{document}
\title{
% Optimality of weighted contracts for multi-agent contract design with a budget 
% Weighted contracts for multi-agent contract design
% Luce contracts
Multi-agent contract design with a budget
% Contract design with multiple independent agents
\thanks{We are grateful to Olivier Bochet, Federico Echenique, Amit Goyal, as well as participants at Winter School of Econometric Society at Delhi School of Economics (2023), EC 2024, Stony Brook International Conference on Game Theory (2024), Annual conference at Indian Statistical Institute Delhi (2024), Durham Economic Theory Conference (2025), and seminar audiences at NYU Abu Dhabi and University of Liverpool for helpful comments and suggestions. An earlier version of this paper circulated under the
title ``Optimality of weighted contracts for multi-agent contract design with a budget." An extended abstract of the manuscript appeared in the proceedings of the 25th
ACM Conference on Economics and Computation (EC’24).}
}

\author{Sumit Goel\thanks{NYU Abu Dhabi; \href{mailto:sumitgoel58@gmail.com}{sumitgoel58@gmail.com}; 0000-0003-3266-9035} \quad Wade Hann-Caruthers\thanks{Technion - Israel Institute of Technology; \href{mailto:whanncar@gmail.com}{whanncar@gmail.com}; 0000-0002-4273-6249}}
% \date{}

\date{\today\\\small{\href{https://goelsumit.com/files/contract_multiagent.pdf}{(Link to latest version)}}}

\maketitle

\begin{abstract}
We study a multi-agent contract design problem with moral hazard. In our model, each agent exerts costly effort towards an individual task at which it may either succeed or fail, and the principal, who wishes to encourage effort, has an exclusive-use budget that it can use to reward the agents. A motivating application is crowdsourcing for innovation, where a fixed budget is provided to a crowdsourcing platform to use for rewarding participants based on their submissions. Our main contribution is to introduce a novel class of contracts, which we call Luce contracts, and show that there is always a Luce contract that is optimal. A (generic) Luce contract assigns weights to the agents and distributes the entire budget among the successful agents in proportion to their weights. Furthermore, we characterize effort profiles that can be implemented by Luce contracts and show that Luce contracts offer a way to mitigate the uncertainty in total payments compared to alternative contracts—such as piece-rate or bonus-pool contracts—suggesting their desirability even in environments without budget constraints.
\end{abstract}

% We study a multi-agent contract design problem with moral hazard. In our model, each agent exerts effort towards a task at which it may independently succeed or fail, as might be the case in a crowdsourcing for innovation environment. Our main contribution is to introduce a novel class of contracts, which we call Luce contracts, and illustrate their desirability in such environments. A (generic) Luce contract assigns weights to the agents and distributes a pre-determined budget among successful agents in proportion to their weights. We show that when the principal has an exclusive-use budget that it can use to reward the agents, there is always a Luce contract that is optimal. In general, we characterize effort profiles that can be implemented by Luce contracts and show that Luce contracts offer a way to mitigate the uncertainty in total payments compared to alternative contracts—such as piece-rate or bonus-pool contracts—suggesting their desirability even in environments without budget constraints.

\section{Introduction}

We study a contract design problem for a principal interacting with many independent agents, as is the case for example in crowdsourcing environments. While the principal could, in theory, treat each agent independently and offer an individually optimal piece-rate contract, this incentive scheme has several undesirable features, including significant variability in total payments and the risk of excessively large payouts. In practice, contracts in such environments are often not piece-rate but instead structured as competitions, where the principal sets aside a fixed budget and commits to distributing the budget among a subset of well-performing agents. For example, crowdsourcing platforms like Wazoku and HeroX invite organizations to pose their unsolved problems as Challenges to a crowd of solvers, outlining criterion for a submission to be successful and specifying a budget for awarding successful submissions. In this paper, we model the contract design problem in the presence of budgetary constraints, and further, formalize the sense in which the competition structure of these contracts helps mitigate the uncertainty in total payments.\\

In our model, each agent exerts costly effort towards a task, in which it may independently succeed or fail, and the principal, who is budget-constrained, can commit to a contract which defines the payment to each agent as a function of the set of agents who are successful. For convenience, we assume that the principal's budget is exclusive-use, meaning that any unused portion of the budget provides no additional value to the principal. The assumption is appropriate when the budget is provided as a restricted-use fund or when agents are rewarded with probability shares for an indivisible prize. It is also reasonable in  environments where the budget is small in comparison to the benefit associated with successful submissions, as might be the case in crowdsourcing for innovation. The principal's problem is then to find a contract, which describes how the budget is distributed among the agents for each outcome, so as to incentivize agents to exert higher effort towards being successful.\\

We first show that, even though there might be multiple optimal contracts, any such contract must be \textit{successful-get-everything} (SGE): it rewards only successful agents and distributes the entire budget among them. In particular, contracts such as piece-rate or bonus-pool contracts, which under certain outcomes distribute only a fraction of the budget to successful agents, can never be optimal in this setting. Intuitively, the principal should be able to improve upon contracts that leave money on the table by using that leftover money to increase the rewards of some successful agent, thus incentivizing them to exert greater effort.\footnote{This intuition is slightly misleading, since equilibrium forces imply that there isn't necessarily an obvious way to distribute this left-over budget and induce higher effort from all agents. Nevertheless, we show that there is always some way to do this, and this implies that any optimal contract must be SGE.} We further show that any SGE contract is optimal for some monotone objective of the principal. In other words, the maximal set of implementable equilibria is exactly the set of equilibria of SGE contracts.  \\

Our main contribution is then to introduce a subclass of SGE contracts, called Luce contracts, and to show that there is always a Luce contract that is optimal. A Luce contract is defined by assigning a weight to each agent and a (weak) priority order over the set of agents, so that the entire budget is distributed among the highest-priority successful agents in proportion to their weights. We show that Luce contracts are minimally sufficient to implement the maximal set, and thus, the principal can always optimize over Luce contracts. This result provides for a significant reduction in the complexity of the design problem, narrowing the search space from a $\Theta(n 2^n)$-dimensional space of complex and difficult-to-interpret contracts to an $(n-1)$-dimensional space of simple and easily interpretable contracts. We apply our results to the special case of two agents with quadratic costs, where our analysis uncovers an interesting robustness in the structure of the optimal contract to the heterogeneity in the agents' cost parameters.\\

Lastly, we discuss the desirability of Luce contracts for implementation in more standard contract design environments, where the principal incurs a direct cost from rewarding the agents but is not budget-constrained. In such environments, the principal has access to a wide range of contracts, including piece-rate and bonus-pool contracts, for the purpose of inducing any desired effort profile. We characterize the set of effort profiles that can be implemented via Luce contracts with some budget and show that, for any such profile, the total payment under any implementing contract is a mean-preserving spread of the total payment under the corresponding Luce contract. It follows that Luce contracts minimize both the maximum total payment the principal might need to make and the variability in total payment for these effort profiles. With these desirable properties, we believe Luce contracts might provide a useful alternative for inducing effort over commonly studied contracts, like piece-rate and bonus-pool contracts, for the purpose of implementation in general multi-agent contract design environments.
 
\subsection*{Related literature}

There is a vast literature on principal-agent problems under moral hazard. In the canonical model with a single agent (\citet*{holmstrom1979moral, grossman1992analysis, mirrlees1976optimal}), the principal offers a wage contract that defines the agent's payment as a function of its observed output, and then the agent chooses some unobserved action (effort) that determines the distribution over outputs. A key finding is that optimal contracts reward the agent for output realizations that are informative about the target level of effort (informativeness principle), and may therefore be non-monotone in output. There has since been significant work studying variants of this single-agent model incorporating flexible actions (\citet*{georgiadis2024flexible}), multiple tasks (\citet*{holmstrom1991multitask, bond2009multitask}), bounded payments (\citet*{jewitt2008moral}), combinatorial actions (\citet*{dutting2022combinatorial}), and informationally robust design (\citet*{carroll2015robustness, zhang2023distributionally}). 
% In particular, \citet*{bond2009multitask} study a related model in which there is a single agent who chooses effort levels for multiple tasks, each of which may succeed or fail, and finds that the optimal contract involves excessive concentration on certain tasks and is fragile. 
See \citet*{georgiadis2022contracting, holmstrom2017pay} and \citet*{lazear2018compensation} for recent surveys of this literature. \\

Our paper contributes to the literature studying a principal contracting with multiple agents. In these settings, agents may be rewarded not only based on their absolute performance but also on their relative performance or even the joint performance of the group. When agents' observed outputs are subject to a common shock or involve subjective evaluations, the optimal contract may involve pay for relative performance (\citet*{green1983comparison, lazear1981rank, malcomson1986rank, malcomson1984work, mookherjee1984optimal, nalebuff1983prizes, imhof2014bonus}). And joint performance evaluation may be optimal in the presence of negatively correlated outputs (\citet*{fleckinger2012correlation}), complementarities in production (\citet*{alchian1972production}), incentives to help  (\citet*{itoh1991incentives}), and unknown production environment (\citet*{kambhampati2024robust}). Other related work on multi-agent contract design explores moral hazard in teams (\citet*{holmstrom1982moral, winter2004incentives, battaglini2006joint, babaioff2012combinatorial, halac2021rank, dai2022robust}), computational challenges in finding optimal contracts (\citet*{castiglioni2023multi,dutting2023multi, ezra2024approximability}), and design problems tailored to more specific environments (\citet*{halac2017contests,  haggiag2022picking, baiman1995informational}). In contrast to this literature, our work focuses on a principal contracting with multiple independent agents, where standard single-agent contracts would be optimal but are rendered infeasible due to budgetary constraints on the principal.\footnote{There is also a related literature in contests, where the contract specifies a distribution of the budget based on how agents rank in terms of their output (\citet*{fang2020turning}). However, in our model, the principal observes only whether agents succeed or fail, not their exact effort levels or rankings. Thus, the reward schemes typically analyzed in contest settings are not applicable to our framework.}

% \subsection*{Literature review}

% \sumit{relationship to contest literature: there, you observe efforts and distribute rewards based on ranks}

% \footnote{\citet*{haggiag2022picking} study a related model, where each agent $i\in [n]$ chooses only whether to participate or not, captured by a choice of $p_i\in \{0, q_i\}$ where $q_i \in (0,1)$ is fixed and common knowledge. They are interested in how different tie-breaking rules influence participation in equilibrium.} 
\section{Model}

There is a principal and $n$ risk-neutral agents. Each agent $i\in [n]$ is assigned an independent task, in which it may succeed or fail. The agents choose how much effort to exert towards succeeding in their respective tasks, captured by their choice of success probabilities: if agent $i$ chooses success probability $p_i\in [0,1]$, it incurs a cost $c_i(p_i)$ in doing so, where $c_i:[0,1]\to \R$ is differentiable, strictly increasing, and strictly convex, with $c_i'(0)=0$. The cost functions are common knowledge.\footnote{We note that our model can also be interpreted in a more general framework but focusing on a restricted class of target-based contracts. Specifically, suppose each agent $i\in [n]$ may choose an effort level $e_i \in \mathbb{R}_+$, which induces a probability distribution over a set of observable outcomes $X_i$. Consider contracts in which the principal specifies a target $t_i\in X_i$ for each agent so that the agent is deemed to have succeeded if the realized outcome $\geq t_i$ and failed otherwise. Under standard assumptions, such as $X_i$ being ordered and higher effort inducing distributions that first-order stochastically dominate those under lower effort, our model can be interpreted as analyzing such target-based contracts within this general setting.}\\

The principal would like to incentivize the agents to maximize the probability that they succeed in their tasks. Formally, we assume that the principal's preference over success probability profiles is represented by a continuous, strictly increasing objective function $V(p_1, \dots, p_n)$. For example, if the principal obtains a profit of $w_i$ if agent $i$ succeeds in its task, and it cares about maximizing expected profit, its preference may be represented by the objective function $$V(p)=\sum_{i=1}^n w_ip_i.$$
Alternatively, if the principal gets a constant positive payoff as long as there is at least one successful agent, and $0$ otherwise (as may be the case in a crowdsourcing for innovation example), its preference may be represented by $$V(p)=1-\Pi_{i=1}^n (1-p_i).$$

In order to incentivize the agents, the principal can design a contract mapping observed outcomes to a reward for each agent. We assume that the principal has an exclusive-use budget $B>0$ that can be used for the purpose of rewarding the agents, and the principal gets no value from unused portions of the budget. Formally, the principal commits to a contract which specifies the fraction of the budget each agent receives depending on the observed outcome.
\begin{definition}
A \textit{contract} is a function $f = (f_1, \dots, f_n) : 2^{[n]} \to \R_+^{[n]}$
such that 
\begin{itemize}
    \item $f_i(S) \geq 0$ (limited liability) and
    \item $\sum_{j \in [n]}{f_j(S)} \leq 1$ (budget constraint)
\end{itemize}
for each $i \in [n]$ and $S \subseteq [n]$.
\end{definition}
Under the contract $f$, if $S$ is the set of agents who succeed, then each agent $i$ receives the reward $f_i(S) \cdot B$. For instance, a natural contract in this environment would be to split the budget equally amongst agents who are successful:
\begin{align*}
    f_i(S) =
    \begin{cases}
        \frac{1}{|S|}, &i \in S \neq \emptyset\\
        0, &\text{otherwise.}
    \end{cases}
\end{align*}
Note that, as in this contract, each agent's reward may depend not only on whether she succeeds, but also on the success of the other agents. We will denote by $\F$ the set of contracts. Throughout the paper, we will work with several subclasses of contracts, and we will in general denote by $\F_{\phi}$ the set of contracts with property $\phi$.\\

A contract $f\in \mathcal{F}$ defines a normal-form game between the $n$ agents, in which each agent chooses $p_i\in [0,1]$ and agent $i's$ payoff under the profile $p=(p_1, \dots, p_n)$ is its expected reward minus the cost of choosing $p_i$. Formally, 
$$u_i(p)= \E_p[f_i(S) \cdot B] - c_i(p_i) = \sum_{S \subset [n]} {\Pr}^{[n]}_p(S) \cdot f_i(S) \cdot B -c_i(p_i)$$ where $${\Pr}^{[n]}_p(S)=\prod_{i \in S} p_i\prod_{j \in [n] \setminus S}(1-p_j).$$

Notice that the expected reward in linear in $p_i$, and since the cost is convex, the utility is strictly concave in $p_i$. It thus follows that pure-strategy Nash equilibrium exists. We will denote by $E : \F \rightrightarrows [0, 1]^n$ the equilibrium correspondence, so $E(f)$ is the set of pure-strategy equilibria for the contract $f$ and $E^{-1}(p)$ is the set of contracts for which $p$ is an equilibrium. We further denote by $\Eq$ the set of implementable profiles:
\begin{align*}
    \Eq := \{p \in [0, 1]^n \, : \, p\in E(f) \text{ for some } f \in \F\}.
\end{align*}

The principal's problem is to choose a contract $f$ that induces an equilibrium $p \in E(f)$ that maximizes its objective $V$. As is standard, we assume that when a contract has multiple equilibria, the principal can select its most preferred equilibrium.\footnote{For an illustration of multiplicity, suppose $B\geq \sum_{i \in [n]} c_i'(1)$ and consider the bonus-pool contract under which agent $i$'s reward is $c_i'(1)$ if all agents succeed and $0$ otherwise. For this contract, it is easy to verify that both $p=(1, 1, \dots, 1)$ and $p=(0, 0, \dots, 0)$ are Nash equilibria. That said, if the budget is sufficiently small and the cost functions are sufficiently convex, we can use the diagonally strict concavity condition of \citet{rosen1965existence} to show that every contract has a unique equilibrium.} Hence, the principal's problem is to maximize its objective $V$ over the set of implementable profiles $\Eq$. As we show, $\Eq$ is compact, and since $V$ is strictly increasing, any maximizer of $V$ must be a maximal element of $\Eq$ with respect to the standard coordinatewise ordering. We denote by $\PO$ the set of maximal equilibrium profiles in $\Eq$. Formally, 
$$\PO = \{p\in \Eq: \forall q\in \Eq, \,\, q_i\geq p_i \,\, \forall i\in[n] \,\, \Rightarrow \,\,  q=p\}.$$
The principal's problem thus reduces to maximizing its objective $V$ over the set of maximal profiles $\PO$:
\begin{align*}
    \max_{p \in \PO}{V(p)}.
\end{align*}

\subsection*{Small budget}

Notice that if the budget is sufficiently large ($B\geq \sum_{i\in [n]} c_i'(1)$), the principal can induce any effort profile $q \in [0, 1]^n$ using the following piece-rate contract:
\begin{align*}
    f_i(S) =
    \begin{cases}
        \frac{1}{B} \cdot c_i'(q_i), &i \in S\\
        0, &\text{otherwise.}
    \end{cases}
\end{align*}
Indeed, agent $i$'s expected utility in this case is $p_i \cdot c_i'(q_i) - c_i(p_i)$ which is uniquely maximized at $p_i = q_i$. It follows that $\Eq=[0,1]^n$, $\PO=\{(1,1,\dots, 1)\}$, and the problem is trivial.\\

Hence, we focus on the more interesting case where the principal's budget is small.  In particular, we assume that the principal's budget is small enough that it could not induce even one of the agents to succeed with probability $1$. 

\begin{assumption}
\label{ass:small-budget}
The principal's budget $B<\min\{c_1'(1), c_2'(1), \dots, c_n'(1)\}$.    
\end{assumption}

Assumption~\ref{ass:small-budget} ensures that in equilibrium, no agent chooses $p_i = 1$. We note that our results extend to situations with a larger budget as long as we restrict attention to equilibria that are interior.

% For convenience, we further normalize the budget to $B = 1$. Note that this is without loss of generality, since for any contract $f\in \F$, the game with budget $B$ and costs $c_i$ is strategically equivalent to the game with budget $1$ and costs $\frac{1}{B} \cdot c_i$.
\section{Main results}

In this section, we identify and characterize a family of contracts whose equilibria coincide precisely with the set of maximal equilibria.\\

Consider any contract $f\in \F$. At any effort profile $p\in [0,1]^n$, agent $i$'s utility can be conveniently expressed as
\begin{equation}
\label{eq:clean}
u_i(p) = \big(\E_p[f_i(S) \given i \in S] \cdot p_i + \E_p[f_i(S) \given i \notin S] \cdot (1-p_i)\big)  \cdot B  - c_i(p_i).
\end{equation} 
Since $u_i$ is strictly concave in $p_i$, agent $i$ has a unique best response characterized by the first-order condition:
\begin{equation}
\label{eq:mainfoc}
c_i'(p_i)=\max\{0, \big(\E_p[f_i(S) \given i \in S] - \E_p[f_i(S) \given i \notin S]\big)\cdot B\}.
\end{equation}
We denote this unique best response by $b_i(f, p_{-i})$. \\

Equation \eqref{eq:mainfoc} makes clear that rewarding agents when they fail reduces their incentive to exert effort, and suggests that the principal should avoid using contracts that reward unsuccessful agents. We introduce this class of contracts formally:

\begin{definition}
A contract $f\in \F$ is \textit{failures-get-nothing} (FGN) if $f_i(S) = 0$ whenever $i \notin S$.
\end{definition}

It turns out that using a contract which is not FGN is wasteful in a very particular sense. Given any such contract and an effort profile $p$ it implements, it is possible to scale the payments down at every outcome to form an FGN contract which also implements $p$. As a consequence, FGN contracts are not only sufficient, but offer the cheapest alternative for inducing any implementable profile.
\begin{proprep}
\label{prop:FGN}
For any implementable profile $p$, there exists an FGN contract $g$ which implements it. Moreover, the expected total payment at $p$ under $g$ is minimal: $$\E_p[ \sum_{i\in [n]} g_i(S)] \cdot B = \sum_{i\in [n]} p_ic_i'(p_i).$$
\end{proprep}
\begin{proof}
Let $p\in \Eq$ and let $f \in E^{-1}(p)$. Consider the contract $g$ defined by
\begin{align*}
    g_i(S)= 
    \begin{cases}
        \lambda_i f_i(S), & \text { if } i \in S \\
        0, & \text { otherwise }
    \end{cases}
\end{align*}
where
\begin{align*}
    \l_i= 
    \begin{cases}
        \dfrac{\E_p[f_i(S) \given i \in S] - \E_p[f_i(S) \given i \notin S]}{\E_p[f_i(S) \given i \in S]} & \text { if } p_i>0 \\ 
        0 & \text { if } p_i=0.
    \end{cases}
\end{align*}
It follows that $\lambda_i\leq 1$ so that $g$ is indeed an FGN contract. Now if $p_i > 0$, agent $i$'s marginal utility at profile $p$ under the contract $g$ is
\begin{align*}
    \dfrac{\partial u_i(p)}{\partial p_i}&= \E_p[g_i(S) \given i \in S] \cdot B - c_i'(p_i) &(\text{Equation } \eqref{eq:clean})\\
    &= \l_i \E_p[f_i(S) \given i \in S] \cdot B - c_i'(p_i)\\
    &= \E_p[f_i(S) \given i \in S]\cdot B - \E_p[f_i(S) \given i \notin S]\cdot B - c_i'(p_i)\\
    &= 0. & (\text{because } p\in E(f))
\end{align*}
By concavity of payoffs, $b_i(g, p_{-i}) = p_i$. If $p_i = 0$, then $g_i(S) = 0$ for all $S$, so $b_i(g, p_{-i}) = 0 = p_i$. Hence, $p\in E(g)$. It is straightforward to verify that the expected total payment under $g$ at $p$ is $\sum_{i=1}^n p_ic_i'(p_i)$, and it is weakly lower than any other $f$ that may induce $p$. 
\end{proof}

Going forward, without loss of generality, we restrict attention to FGN contracts. Under any FGN contract $f$ and effort profile $p$, Equation \eqref{eq:mainfoc} simplifies to:
\begin{equation}
\label{eq:foc_fgn}
c_i'(p_i) = \mathbb{E}_p[f_i(S) \mid i \in S] \cdot B.
\end{equation}

This equation shows clearly that an agent's incentive to exert effort increases in their expected share of the budget if they succeed. Since the principal's budget is exclusive-use, meaning that it does not derive any value from leftover budget, this suggests that the principal should always distribute the entire budget among successful agents.

\begin{definition}
A contract $f$ is \textit{successful-get-everything} (SGE) if 
\begin{itemize}
    \item $f_i(S) = 0$ whenever $i \notin S$ and 
    \item $\sum_{i \in S}{f_i(S)} = 1$ whenever $S \neq \emptyset$. 
\end{itemize}
\end{definition}

Given any implementable profile $p$, if the FGN contract $f$ which induces it is not SGE, there must be some leftover budget under some outcomes. Observe that the most the principal can give in expected rewards at $p$ is $\P_p[S \neq \emptyset] \cdot B$, and the actual amount it gives in expected rewards is $\E_p[\sum_i f_i(S)]\cdot B$. We denote by $z(p)$ this gap between the upper bound and the actual amount given in rewards, and note that it depends only on $p$:
\begin{align*}
    z(p) &= \P_p[S \neq \emptyset] \cdot B - \E_p\left[\sum_i f_i(S)\right]\cdot B\\
    &= \left[1 - \left(\prod_{i\in [n]} (1-p_i)\right) \right] \cdot B - \sum_{i\in [n]} p_ic_i'(p_i),
\end{align*}
where the second term follows from Proposition \ref{prop:FGN}. For any $p\in \Eq$, $z(p)$ provides a useful robust measure of the leftover budget from inducing $p$ using any FGN contract $f\in E^{-1}(p)$.

\begin{corollary}
\label{cor:eq_sge}
If $p$ is implementable, then $z(p) \geq 0$. Moreover, if $p$ is implementable by an SGE contract, then $z(p) = 0$.
% If $p\in E(f)$ and $f\in \F_{SGE}$, then $z(p)= 0$.
\end{corollary}

Now suppose effort profile $p$ is an equilibrium of a SGE contract. If $q$ is an effort profile that coordinate-wise dominates $p$, we can show that $z(q)<z(p)=0$, and so by Corollary \ref{cor:eq_sge}, $q$ is not implementable. It thus follows that any equilibrium of any SGE contract is maximal. \\
% \wade{Is this the right amount of detail?} \\

Interestingly, maximal equilibria can only be induced by SGE contracts.  Intuitively, if $p$ is an an equilibrium of a FGN contract which is not SGE, the leftover budget could potentially be used to incentivize higher effort. 
% In particular, there must be some outcome $S$ where the total reward is less than the budget, $\sum_i f_i(S) \cdot B < B$.  
While one might suspect that increasing agent $i$'s share $f_i(S)$ for some $i\in S$ (if $\sum_{i\in S} f_i(S)<1$) might do the trick, the equilibrium forces imply that such a transformation does not necessarily lead to a dominant outcome.\footnote{For an illustration, consider an example with $n=2$, $c_i(p_i)=\frac{1}{2}C_ip_i^2$, and $B=1$. Suppose $f\in \F$ is such that it awards $\beta_i\in [0,1]$ to agent $i$ if and only if $i$ is the only successful agent. Using Equation \eqref{eq:foc_fgn}, one can verify that $f$ has a unique equilibrium $p=(p_1, p_2)$ with 
$p_i=\dfrac{\beta_i(C_{-i}-\beta_{-i})}{C_1C_2-\beta_1\beta_2}.$ From here, it is easy to check that increasing $\beta_i$ would increase $p_i$, but it may also be accompanied by a decrease in $p_{-i}$.} In general, even though there isn't an obvious way to transform these contracts so that they induce dominant outcomes, we show there always exists some way.

\begin{thmrep}
\label{thm:pareto-optimal-iff-sge}
Suppose the contract $f$ implements the effort profile $p$. Then, $p$ is maximal if and only if $f$ is an SGE contract.
\end{thmrep}
\begin{proof}
Suppose $f$ is a SGE contract and $p\in E(f)$. We want to show that $p$ is maximal. Observe that
\begin{align*}
    \dfrac{\partial z}{\partial x_i}\Bigr|_{x\geq p}&=B \prod_{j\neq i} (1-x_j) -x_ic_i''(x_i) - c_i'(x_i)\\
    &< B\prod_{j\neq i} (1-x_j) - c_i'(x_i) &(\text{because } c_i \text{ is convex})\\
    &\leq B\prod_{j\neq i} (1-x_j) - c_i'(p_i) &(\text{because } c_i \text{ is convex})\\
    &\leq B \prod_{j\neq i} (1-p_j) - c_i'(p_i) &(\text{because } x_j\geq p_j)\\
    &\leq B \cdot \E_p[f_i(S)|i\in S]  - c_i'(p_i) &(\text{because } f\in \F_{SGE})\\
    &= 0 &(\text{because }p\in E(f))
\end{align*}

It follows that if effort profile $q$ dominates $p$, then $z(q)<z(p)$, and from Corollary \ref{cor:eq_sge}, $q$ is not implementable. Thus, $p$ is maximal. \\

Now suppose $p$ is maximal. We want to show that any $f\in E^{-1}(p)$ must be SGE. We first show that all agents must be active at $p$.

\begin{lem}
\label{lem:no-zero}
For all $i\in [n]$, $0<p_i<1$. 
\end{lem}
\begin{nestedproof}
Suppose towards a contradiction that $p_k = 0$ for some $k\in [n]$. Let $f\in E^{-1}(p)\cap \F_{FGN}$ and consider the contract $g$ where $g_i(S) = f_i(S \setminus \{k\})$ for $i \neq k$ and
$$
g_k(S)= \begin{cases}1, & \text { if } S = \{k\} \\ 0, & \text { otherwise }\end{cases}
$$
It is easy to verify that $p' = (p_k', p_{-k})$ where $p_k'=b_k(g, p_{-k})>0$ is an equilibrium under $g$. But then, $p'\in \Eq$ dominates $p$, which contradicts the fact that $p$ is maximal. 
\end{nestedproof}

Now let
\begin{align*}
    K_p := \{ S \subseteq [n] \, : \, \sum_{i \in S}{f_i(S)} < 1 \text{ for some } f \in E^{-1}(p) \cap \mathcal{F}_{FGN}\},
\end{align*}
and let
\begin{align*}
    \kappa_p := \{ i \in [n] \, : \, \{i\} \in K_p \}.
\end{align*}

We now show that $K_p = \{\emptyset\}$. 
\begin{enumerate}

\item Step 1: Suppose $S \in K_p$. For any $T \subset S$, $T \in K_p$.  

Let $f \in E^{-1}(p) \cap \mathcal{F}_{FGN}$ be such that $\sum_{i \in S}{f_i(S)} < 1$. If $\sum_{i \in T}{f_i(T)} < 1$, we are done. Otherwise, pick an agent $i\in T$ such that $f_i(T)>0$ and consider a contract $g$ which differs from $f$ only in its award for agent $i$ at $S$ and $T$. In particular, let $g$ be such that $g_i(S)=f_i(S)+\epsilon$ and $g_i(T)=f_i(T)-\delta$ where $\epsilon, \delta>0$ are chosen so that $p \in E(g)$. We can do this because we know from Lemma \ref{lem:no-zero} that all events occur with positive probability at profile $p$. It follows then that $g \in E^{-1}(p) \cap \mathcal{F}_{FGN}$ and $\sum_{i \in T}{g_i(T)} < 1$. Thus, $T\in K_p$.

\item Step 2: Suppose $S, T \in K_p$. Then,  $S \cup T \in K_p$.  

Let $f,g \in E^{-1}(p) \cap \mathcal{F}_{FGN}$ be such that $\sum_{i \in S}{f_i(S)} < 1$ and $\sum_{i \in T}{g_i(T)} < 1$. Consider the contract $h=\frac{1}{2}f+\frac{1}{2}g$. From  Equation~\eqref{eq:foc_fgn}, $h \in E^{-1}(p) \cap \mathcal{F}_{FGN}$ and also $\sum_{i \in S}{h_i(S)} < 1$ and $\sum_{i \in T}{h_i(T)} < 1$. Now, if $\sum_{i \in S \cup T}{h_i(S \cup T)} < 1$, we are done. Otherwise, pick any agent $i\in S \cup T$ (WLOG, let $i\in S$) such that $h_i(S \cup T)>0$ and consider a contract $h'$ which differs from $h$ only in its award for agent $i$ at $S\cup T$ and $S$. In particular, let $h'$ be such that $h'_i(S\cup T)=h_i(S \cup T)-\epsilon$ and $h'_i(S)=h_i(S)+\delta$ where $\epsilon, \delta>0$ are chosen so that $p \in E(h')$. It follows then that $h' \in E^{-1}(p) \cap \mathcal{F}_{FGN}$ and $\sum_{i \in S \cup T}{h'_i(S\cup T)} < 1$. Thus, $S \cup T\in K_p$.

Note that it follows from Steps 1 and 2 that $K_p = 2^{\kappa_p}$.

\item Step 3:  Suppose $f\in E^{-1}(p) \cap \mathcal{F}_{FGN}$. Then, for all $S \subset [n]$ such that $\kappa_p^C \cap S \neq \phi$, $f_i(S)=0$ for all $i \in \kappa_p$.

Suppose towards a contradiction that there is an $S \subset [n]$ such that  $\kappa_p^C \cap S \neq \phi$ and $f_i(S)>0$ for some $i \in \kappa_p$. Let $g \in E^{-1}(p) \cap \mathcal{F}_{FGN}$ be such that $g_i(\{i\}) < 1$. Consider the contract $h=\frac{1}{2}f+\frac{1}{2}g$. Then $h \in E^{-1}(p) \cap \mathcal{F}_{FGN}$ and also $h_i(S)> 0$ and $h_i(\{i\}) < 1$. Now, consider a contract $h'$ which differs from $h$ only in its award for agent $i$ at $S$ and $\{i\}$. In particular, let $h'$ be such that $h'_i(\{i\})=h_i(\{i\})+\epsilon$ and $h'_i(S)=h_i(S)-\delta$ where $\epsilon, \delta>0$ are chosen so that $p \in E(h')$. It follows then that $h' \in E^{-1}(p) \cap \mathcal{F}_{FGN}$ and $\sum_{i \in S}{h'_i(S)} < 1$. But this means that $S \subset \kappa_p$ which is a contradiction.

\item Step 4: Suppose $\kappa_p \neq \phi$. Then there is a $p'\in \Eq$ that dominates $p$.

For all $S\in K_p$, let $f^S \in E^{-1}(p) \cap \mathcal{F}_{FGN}$ be such that $\sum_{i \in S}{f^S_i(S)} < 1$. Consider the contract $g=\sum_{S \in K_p} \frac{1}{|K_p|}f^S$. Then $g \in E^{-1}(p) \cap \mathcal{F}_{FGN}$ and also $\sum_{i \in S}{g_i(S)} < 1$ for all $S\in K_p$. Now, we can construct a contract $h \in E^{-1}(p) \cap \mathcal{F}_{FGN}$ such that
$$
h_i(S)= \begin{cases}g_i(S), & \text { if } S \in K_p\\ h_i(S \setminus \kappa_p), & \text { if } S \notin K_p\end{cases}
$$
by averaging over the outcomes of agents in $\kappa_p$ under $g$.

% In words, the reward for agent $i\notin \kappa_p$ does not depend on the outcomes of agents in $\kappa_p$. Now $p\in E(h)$ because from the perspective of agents in $\kappa_p$, $h$ is the same as $g$, and thus, their best response to $p_{-i}$ is still $p_i$. For agent $i \notin \kappa_p$, the averaging is such that 
% $\sum_{S\subset [n]_{-i}} g_i(S \cup \{i\}){\Pr}^{[n]_{-i}}_{p_{-i}}(S)=\sum_{S\subset [n]_{-i}} h_i(S \cup \{i\}){\Pr}^{[n]_{-i}}_{p_{-i}}(S)$ and thus, $p_i$ continues to be the unique best response to $p_{-i}$ under $h$. 

% Thus, we have $h \in E^{-1}(p) \cap \mathcal{F}_{FGN}$ and also $\sum_{i \in S}{h_i(S)} < 1$ for all $S\in K_p$. 

Observe that if we manipulate $h$ at any $S\subset \kappa_p$, it won't change the best responses for agents $i\notin \kappa_p$. We will now show that we can manipulate the awards for $S\subset \kappa_p$ so that under the new contract $h'$,  $p'\in E(h')$ where $p'_i>p_i$ for $i\in \kappa_p$ and $p'_i=p_i$ for $i\notin \kappa_p$. Towards this goal, let $A=\kappa_p$ and let $p'=(p_i+\epsilon)_{i \in A}$. For each $i\in A$,  let $t_i(\epsilon)$ solve

$$c_i'(p'_i)= B \cdot t_i(\epsilon) \cdot \sum_{S\subset A_{-i}} (h_i(S \cup \{i\})){\Pr}^{A_{-i}}_{p'_{-i}}(S)$$

Observe that as $\epsilon\to 0$, $t_i(\epsilon)\to 1$ for all $i\in A$. Since $\sum_{i \in S}{h_i(S)} < 1$ for all $S\subset A$ and $t_i(\epsilon)$ is continuous in $\epsilon$, we can find $\epsilon>0$ small enough so that the contract $h'_i(S)=h_i(S)*t_i(\epsilon)$ for all $S\subset A $ and $i\in S$ is a feasible contract. By construction, the profile $(p', p_{-A})$ will be an equilibrium under $h'$. Thus, we have that $p$ is not maximal.

\end{enumerate}
It follows then that $K_p=\{\emptyset\}$. By definition of $K_p$, this means that for any $f\in E^{-1}(p)$, either $f\notin \mathcal{F}_{FGN}$ or $f\in \mathcal{F}_{SGE}$. Now suppose there exists an $f$ such that $f\notin \mathcal{F}_{FGN}$ and $p\in E(f)$. From Proposition \ref{prop:FGN}, we can find $g\in \mathcal{F}_{FGN}$ such that $p\in E(g)$. Moreover, we know from the construction of $g$ in the argument of Proposition \ref{prop:FGN} that $g\notin \mathcal{F}_{SGE}$. Thus, we have that $g\in \mathcal{F}_{FGN}, g\notin \mathcal{F}_{SGE},$ and $p\in E(g)$. But this means that $K_p\neq \emptyset$ which is a contradiction. 
Therefore, it must be that for any $f\in E^{-1}(p)$,  $f\in \mathcal{F}_{SGE}$.\\

\end{proof}

As a consequence, since contracts like piece-rate (where each agent $i$ gets a fixed reward if she is successful and $0$ otherwise) or bonus-pool (where each agent gets a nonzero share of the budget only if all agents succeed) are not SGE, they never induce maximal equilibria. In comparison, SGE contracts induce competition between the agents as they compete for a fixed budget, and always lead to maximal equilibria. Thus, the result suggests that in environments where the principal operates with an exclusive-use budget, fostering competition among the agents through successful-get-everything contracts creates stronger incentives than promoting teamwork through bonus-pool contracts or independent performance assessment through piece-rate contracts.\\

While these simpler, well-known contracts should be avoided in favor of SGE contracts, the set of all SGE contracts itself lacks clear structure, being $\Theta(n 2^n)$-dimensional. Consequently, optimizing over SGE contracts may still pose a computationally difficult problem, and the resulting optimal contract may be complex to interpret and implement in practice. Now, the set of maximal equilibria lies on the boundary of the equilibrium set and is thus at most $(n-1)$-dimensional. This dimensionality gap suggests that for any given maximal equilibrium, there are potentially many SGE contracts that implement it, which raises the question: is there a simpler, low-dimensional, structured subclass of SGE contracts that the principal can use? Weighted contracts, which are both structured and dimensionally aligned, provide a natural and promising candidate.

% While the principal should not use these simple and well-studied contracts, and should restrict attention to the set of SGE contracts, this set does not have much structure and is $\Theta(n 2^n)$-dimensional. As a consequence, optimizing over SGE contracts may still present the principal with a computationally difficult problem, and even an optimal contract may be difficult to understand and implement. Now, the set of maximal equilibria is contained in the boundary of the set of equilibria, so it is at most $n-1$-dimensional, which suggests that for any given maximal equilibrium, there are potentially many SGE contracts that implement it. This presents the question: is there a simple, low-dimensional, structured set of contracts that the principal can use? Weighted contracts, which are highly structured and have matching dimensionality, provide a natural candidate.

\begin{definition}
A contract $f$ is a \textit{weighted} (W) contract if there exist weights $(\l_1, \dots, \l_n)$ with $\l_i > 0$ such that 
\begin{align*}
    f_i(S) = 
    \begin{cases}
        \dfrac{\l_i}{\sum_{j \in S}{\l_j}}, &\text{if} \, i \in S\\
        0, &\text{otherwise}
    \end{cases}
\end{align*}
\end{definition}

A weighted contract assigns weights to the agents, and distributes the entire budget among successful agents in proportion to their weights. As the following result shows, these simple contracts do in fact essentially span the whole set of maximal equilibria.

\begin{thmrep}
\label{thm:weighted}
For every maximal equilibrium $p$, there is a unique contract in the closure of weighted contracts which implements $p$.    
\end{thmrep}

\begin{proof}
From Theorem \ref{thm:pareto-optimal-iff-sge}, we know that $p$ must be an equilibrium of a SGE contract. But then, it follows that $p$ must satisfy \eqref{eq:condition}. From Proposition~\ref{prop:necc-suff-pw}, there is a Luce contract that implements $p$. Thus, we get that every maximal equilibrium can be implemented by a Luce contract. \\

Now we show that there is a unique Luce contract that implements $p$. Consider the case where $f,g$ are weighted contracts and suppose towards a contradiction that $p\in E(f)\cap E(g)$. Let $i\in [n]$ denote the agent with the smallest weight ratio $\frac{\lambda^g_i}{\lambda_i^f}$, where $\lambda^f, \lambda^g$ represent the weights that define the contracts $f, g \in \mathcal{F}_{W}$. It follows from the definition of weighted contracts that for any $S\subset [n]$ such that $i\in S$, agent $i$'s reward is weakly lower under $g$ than $f$. As a result, it can't be that $b_i(g,p_{-i})=b_i(f,p_{-i})$. It follows that two different weighted contracts cannot have the same equilibrium. \\

Now suppose $f,g$ are arbitrary Luce contracts that implement $p$. If $\succcurlyeq_f = \succcurlyeq_g$, then the same argument works. So suppose that $\succcurlyeq_f \neq \succcurlyeq_g$. Then there is some $i$ such that $T = \{k : k \succcurlyeq_f i\} \neq \{k : k \succcurlyeq_g i\} = T'$, and we may assume without loss of generality that $T' \not \subseteq T$ by exchanging $f$ and $g$ if necessary. Observe that for any $S$,
\begin{align*}
    \sum_{k \in T} g_k(S) \leq \indic_{S \cap T \neq \emptyset} = \sum_{k \in T} f_k(S).
\end{align*}
Moreover, for any $j \in T' \setminus T$,
\begin{align*}
    \sum_{k \in T} g_k(\{i, j\}) = g_i(\{i, j\}) < 1 = f_i(\{i, j\}) = \sum_{k \in T} f_k(\{i, j\}).
\end{align*}
Thus,
\begin{align*}
    \sum_{k \in T} \P[k \in S] \cdot \E[g_k(S) \given k \in S] = \E[\sum_{k \in T} g_k(S)] < \E[\sum_{k \in T} f_k(S)] = \sum_{k \in T} \P[k \in S] \cdot \E[f_k(S) \given k \in S].
\end{align*}
However, since $p$ is an equilibrium for $f$ and $g$, both of the outside expressions must be equal to $\frac{\sum_{k \in T} p_k \cdot c_k'(p_k)}{B}$, contradiction.
\end{proof}

The closure of weighted contracts  also has a relatively simple structure. We refer to these contracts, containing weighted contracts and their limit points, as Luce contracts.

\begin{definition}
A contract $f$ is a \textit{Luce contract} if there exist weights $(\l_1, \dots, \l_n)$ with $\l_i > 0$ and a non-strict ordering $\priority$ on the agents such that
\begin{align*}
    f_i(S) = 
    \begin{cases}
        \dfrac{\l_i}{\sum_{j \in \Top(S)}{\l_j}}, &\text{if} \, i \in \Top(S)\\
        0, &\text{otherwise}
    \end{cases}
\end{align*}
where $\Top(S) = \{i \in S \, : \, i \priority j \, \, \forall j \in S\}$.    
\end{definition}
Luce contracts reward each highest-priority successful agent with a fraction of the budget proportional to her weight.\footnote{The choice of terminology is motivated by the analogy between
the structure of Luce contracts and the selection rules that satisfy Luce’s choice axiom (\citet*{luce1959individual}). Just as Luce rules have the property that the relative odds of selecting one item over another from a pool is unaffected by which other items are in the pool, Luce contracts
have the property that the reward of a successful agent relative to another successful agent is
independent of which other agents succeed.}

\begin{lemrep}
\label{lem:luce=weighted}
A contract $f$ is in the closure of weighted contracts if and only if it is a Luce contract.
\end{lemrep}

\begin{proof}

To begin, suppose that $f$ is the limit of a sequence $g^m$ of weighted contracts with weights $(\l_1^m, \dots, \l_n^m)$. Define the relation $\succeq$ by $i \succeq j$ if and only if $f_i(\{i, j\}) > 0$. It is immediate that $\succeq$ is complete. Moreover, since $f_i(\{i, j\}) > 0$ if and only if there exists an $\eps > 0$ such that $\l_i^m \geq \eps \l_j^m$ for all $m$, $\succeq$ is also transitive, and hence it is a non-strict order. For each $i$, let $S^i = \{j \, : \, i \sim j\}$, and define $\l_i = f_i(S^i)$. We claim that $f$ is a Luce contract with priority ordering given by $\succeq$ and weights given by $(\l_i)$. To prove this, fix $i$ and $S$.\\

\noindent If $i \notin S$, then
\begin{align*}
    f_i(S) = \lim_{m \to \infty} g^m_i(S) = \lim_{m \to \infty} 0 = 0.
\end{align*}
If $i \in S \setminus Top_{\succeq}(S)$, then there is some $j \in S$ such that $j \succ i$. Now,
\begin{align*}
    g^m_i(S) = \frac{\l_i^m}{\sum_{k \in S} \l_k^m} \leq \frac{\l_i^m}{\l_i^m + \l_j^m} = g^m_i(\{i, j\}),
\end{align*}
so
\begin{align*}
    f_i(S) = \lim_{m \to \infty} g^m_i(S) \leq \lim_{m \to \infty} g^m_i(\{i, j\}) = f_i(\{i, j\}) = 0.
\end{align*}
Finally, observe that for any $i$ and $j$, if $i \succ j$, then
\begin{align*}
    \lim_{m \to \infty} \frac{1}{1 + \frac{\l_j^m}{\l_i^m}} = \lim_{m \to \infty} \frac{\l_i^m}{\l^i_m + \l_j^m} = \lim_{m \to \infty} g^m_i(\{i, j\}) = f_i(\{i, j\}) = 1,
\end{align*}
so $\lim_{m \to \infty} \frac{\l_j^m}{\l_i^m} = 0$, and if $i \sim j$, then $S^i = S^j$, so
\begin{align*}
    \lim_{m \to \infty} \frac{\l_j^m}{\l_i^m} = \lim_{m \to \infty} \frac{\frac{\l_j^m}{\sum_{k \in S^j} \l_k^m}}{\frac{\l_i^m}{\sum_{k \in S^i} \l_k^m}} = \lim_{m \to \infty} \frac{g^m_j(S^j)}{g^m_i(S^i)} = \frac{f_j(S^j)}{f_i(S^i)} = \frac{\l_j}{\l_i} \cdot
\end{align*}
Hence, if $i \in Top_{\succeq}(S)$, then
\begin{align*}
    f_i(S) &= \lim_{m \to \infty} g^m_i(S) = \lim_{m \to \infty} \frac{\l^m_i}{\sum_{k \in S} \l^m_k} = \lim_{m \to \infty} \frac{1}{\sum_{k \in Top_{\succeq}(S)} \frac{\l^m_k}{\l^m_i} + \sum_{k \in S \setminus Top_{\succeq}(S)} \frac{\l^m_k}{\l^m_i}}\\
    &= \frac{1}{\sum_{k \in Top_{\succeq}(S)} \frac{\l_k}{\l_i}} = \frac{\l_i}{\sum_{k \in Top_{\succeq}(S)} \l_k} \cdot
\end{align*}
Since $f$ was arbitrary, it follows that $\bar{F}_W \subseteq \F_{Luce}$.\\

Now, suppose $f$ is a Luce contract. For each $i$, let $r_i = |\{j \, : \, j \succeq i\}|$, and for every $t > 0$, let $g^t$ be the weighted contract with weights given by $\l^t_i = t^{r_i} \cdot \l_i$. Now, fix $i$ and $S$. If $i \notin S$,
\begin{align*}
    \lim_{t \to 0} g^t_i(S) = \lim_{t \to 0} 0 = 0 = f_i(S).
\end{align*}
If $i \in S \setminus Top_{\succeq}(S)$, then there is some $j \in S$ such that $j \succ i$ and hence $r_j < r_i$. In particular,
\begin{align*}
    \lim_{t \to 0} g^t_i(S) = \lim_{t \to 0} \frac{t^{r_i} \cdot \l_i}{\sum_{k \in S} t^{r_k} \cdot \l_k} \leq \lim_{t \to 0} \frac{t^{r_i} \l_i}{t^{r_j} \l_j} = 0 = f_i (S).
\end{align*}
Finally, if $i \in Top_{\succeq}(S)$, then
\begin{align*}
    g^t_i(S) = \frac{t^{r_i} \l_i}{\sum_{k \in S} t^{r_k} \cdot \l_k} &= \frac{t^{r_i} \l_i}{t^{r_i} \cdot \sum_{k \in Top_{\succeq}(S)} \l_k + \sum_{k \in S \setminus Top_{\succeq}(S)} t^{r_k} \cdot \l_k}\\
    &= \frac{\l_i}{\sum_{k \in Top_{\succeq}(S)} \l_k + \sum_{k \in S \setminus Top_{\succeq}(S)} t^{r_k - r_i} \cdot \l_k}.
\end{align*}
Since $r_k > r_i$ for all $k \in S \setminus Top_{\succeq}(S)$, it follows that
\begin{align*}
    \lim_{t \to 0} g^t_i(S) = \frac{\l_i}{\sum_{k \in Top_{\succeq}(S)} \l_k} = f_i(S).
\end{align*}
Thus, $f = \lim_{t \to 0} g^t$. Since $f$ was arbitrary, it follows that $\F_{Luce} \subseteq \bar{\F}_W$.
\end{proof}

It follows that the principal can always just optimize its objective over the class of Luce contracts.

\begin{corollary}
\label{cor:luce}
For every maximal equilibrium $p$, there is a unique Luce contract which implements $p$. Thus, for any strictly increasing continuous objective $V(p)$, $$\sup_{f\in \mathcal{F}} \sup_{p \in E(f)} V(p)= \max_{f\in \mathcal{F}_{Luce}} \max_{p \in E(f)} V(p).$$
\end{corollary}

Thus, Theorem~\ref{thm:weighted} provides a significant reduction in the complexity of the principal's optimization problem, reducing the search space from a $(2^{n-1}(n-2) + 1)$-dimensional space of difficult to interpret contracts to an $(n-1)$-dimensional space of easily interpretable contracts.\\

Despite this reduction in complexity, it remains difficult to explicitly solve for the optimal Luce contract for a given objective in general. We are able to solve the problem for the special case where there are two agents with quadratic costs.

\begin{proprep}
\label{prop:example}
Suppose $n = 2$, $c_i(p_i) = \frac12 C_i p_i^2$ with $C_i>B$, and $V(p_1, p_2) = w p_1 +  \cdot p_2$. Then, the optimal contract, defined by $\lambda_1(w)$, takes the form
$$
f_i(S)= \begin{cases}0, & \text { if } i\notin S\\ 
1, & \text { if } S=\{i\}\\
\l_i(w),  & \text { if } S=\{1, 2\}
\end{cases},
$$
where $\l_2(w)=1-\l_1(w)$. Moreover, $\l_1(w)$ is increasing in $w$ and in particular,

$$
\l_1(w)= \begin{cases}0, & \text { if } w \leq \frac{C_1 C_2 - BC_1}{C_1 C_2 + BC_2 - B^2} \\ 
\frac{1}{2}, & \text { if } w=1\\
1,  & \text { if } w\geq \frac{C_1 C_2 + BC_1 - B^2}{C_1 C_2 - BC_2}
\end{cases}.
$$
\end{proprep}
\begin{proof}
We know from Theorem \ref{thm:pareto-optimal-iff-sge} that we can restrict attention to SGE contracts. And with only two agents, the set of SGE contracts can be parametrized by a single parameter $\l$, where $f_i(S) = 0$ whenever $i \notin S$, $f_i(S) = 1$ whenever $S = \{i\}$, and $f_1(\{1, 2\}) = 1 - f_2(\{1, 2\}) = \l$.  The equilibrium conditions for this contract from Equation \eqref{eq:foc_fgn} are
\begin{align*}
    C_1p_1 &= (1 - p_2 + \l p_2) \cdot B\\
    C_2p_2 &= (1 - p_1 + (1-\l)p_1) \cdot B.
\end{align*}
Going forward, we normalize $C_i=\frac{C_i}{B}$. For each $\l$, this system of equations has a unique solution:
\begin{equation}
\label{eqbm}
    p_1(\l)=\frac{C_2-(1-\lambda)}{ C_1C_2-\lambda(1-\lambda)} \quad p_2(\l)=\dfrac{C_1-\lambda}{C_1C_2-\lambda(1-\lambda)} \cdot
\end{equation}
Hence, the principal's problem is equivalent to:
\begin{align*}
    \max_{\l \in [0,1]}{ wp_1(\l)+ p_2(\l)}.
\end{align*}

Using the first order condition, this is maximized either at $\l = 0$, or $\l = 1$, or where
\begin{align*}
    \frac{p_2'(\l)}{p_1'(\l)} = -w \cdot
\end{align*}

Note that
\begin{align*}
    p_1'(\l) = \frac{(C_1 C_2 - \l (1 - \l)) - (C_2 - (1 - \l))(2\l-1))}{(C_1 C_2 - \l (1 - \l))^2} = \frac{C_1C_2 - C_2(2\l-1) - (1 - \l)^2}{(C_1 C_2 - \l (1 - \l))^2}
\end{align*}
and
\begin{align*}
    p_2'(\l) = \frac{-(C_1 C_2 - \l (1 - \l)) - (C_1 - \l)(2\l-1)}{(C_1 C_2 - \l (1 - \l))^2} = \frac{-C_1C_2 - C_1 (2\l-1) + \l^2}{(C_1 C_2 - \l (1 - \l))^2}
\end{align*}
so
\begin{align*}
    \frac{p_2'(\l)}{p_1'(\l)} = \frac{-C_1C_2 - C_1 (2\l-1) + \l^2}{C_1C_2 - C_2(2\l-1) - (1 - \l)^2} = -\frac{C_1}{C_2} \cdot \frac{- \l^2/C_1 + 2\l + C_2 - 1}{-(1-\l)^2/C_2 + 2(1-\l) + C_1 - 1}
\end{align*}
Now observe that the numerator is increasing for $\l < C_1$ and the denominator is decreasing for $\l > -(C_2-1)$. In particular, the fraction is monotonically strictly increasing for $0 < \l < 1$, so $\frac{p_2'(\l)}{p_1'(\l)}$ is monotonically strictly decreasing. It follows then that there is a function $\l_1(w)$ such that the unique optimal choice of $\l$ is $\l_1\left( w \right)$ and it is increasing in $w$. In particular, 
\begin{align*}
    \frac{p_2'(\l)}{p_1'(\l)} \leq \frac{p_2'(0)}{p_1'(0)} = -\frac{C_1}{C_2} \cdot \frac{C_2 - 1}{-1/C_2 + C_1 + 1} = -\frac{C_1 C_2 - C_1}{C_1 C_2 + C_2 - 1}
\end{align*}
and
\begin{align*}
    \frac{p_2'(\l)}{p_1'(\l)} \geq \frac{p_2'(1)}{p_1'(1)} = -\frac{C_1}{C_2} \cdot \frac{- 1/C_1 + C_2 + 1}{C_1 - 1} = -\frac{C_1 C_2 + C_1 - 1}{C_1 C_2 - C_2}
\end{align*}
Now if $w\leq -\frac{p_2'(0)}{p_1'(0)} $, the objective is decreasing in $\l$ and thus $\l_1(w)=0$. And if $w\geq -\frac{p_2'(1)}{p_1'(1)} $, the objective is increasing in $\l$ and thus $\l_1(w)=1$.

Lastly, observe that 
\begin{align*}
    \frac{p_2'(\frac12)}{p_1'(\frac12)} = -1
\end{align*}
irrespective of the costs $C_1,C_2$. And thus, if $w_1=w_2$, we get that $\l^*=\frac{1}{2}$ no matter how heterogeneous the agents are.

\end{proof}

Proposition~\ref{prop:example} shows that an agent's share of the budget is increasing in the relative weight that the principal places on that agent's success. Perhaps more surprisingly, the result also prescribes an equal distribution of the budget whenever the principal values agents' successes equally ($\lambda_1(1) = \frac{1}{2}$), regardless of heterogeneity in their cost parameters $C_1, C_2$. Moreover, because $\lambda_1(w)$ is increasing, the optimal contract always awards a greater share of the reward to the agent whose success the principal values more, independently of the agents' costs. This robustness of the structure of optimal contract to cost heterogeneity is intriguing. An interesting open question is whether this robustness property holds more broadly, and if so, under what general conditions.
\section{Luce contracts without budget constraints}

In the previous section, we established that Luce contracts are optimal in environments where the principal has access to an exclusive-use budget. However, this assumption departs from standard contract-design models, where rewarding agents entails a direct cost, and the principal must weigh this cost against the benefits of inducing higher effort. In this section, we argue that Luce contracts remain desirable even when the principal is not budget-constrained.\\

In the absence of a budget constraint, the principal has access to a broad range of contracts for implementing any desired effort profile $q\in (0,1)^n$. For instance, the principal could offer each agent $i\in [n]$ an independent piece-rate contract that pays $c_i'(q_i)$ upon success and nothing otherwise. The principal could also offer a bonus-pool contract which pays $\dfrac{q_ic_i'(q_i)}{\Pi_{i \in [n]} q_i}$ to agent $i\in [n]$ if all agents succeed, and nothing otherwise. Perhaps, the principal could also just set aside a budget $B=\frac{\sum_{i\in [n]} q_i \cdot c_i'(q_i)}{\P_q[S \neq \emptyset]}$, and design a Luce contract that implements $q$. Unfortunately, this may not always be feasible, as the structure of Luce contracts introduces constraints on effort profiles that they can implement. The following result characterizes effort profiles for which implementation via a Luce contract is feasible.

\begin{proprep}
\label{prop:necc-suff-pw}
Suppose $p\in (0,1)^n$. There exists a Luce contract $f\in \F_{Luce}$ with budget $B>0$ that implements $p$ if and only if for all $I\subset [n]$, 
\begin{equation}
\label{eq:condition}
    \sum_{i\in I} p_i \cdot c_i'(p_i) \leq \P_p[S\cap I\neq \emptyset] \cdot B,
\end{equation}
where
$B=\dfrac{\sum_{i\in [n]} p_i \cdot c_i'(p_i)}{\P_p[S \neq \emptyset]}$.
\end{proprep}

\begin{proof}
Suppose $f\in \F_{Luce}$ with budget $B>0$ implements $p$. Then for any $I \subseteq [n]$, the expected total reward to agents in $I$ is
\begin{align*}
    \sum_{i\in I} p_i \cdot c_i'(p_i) &=  \E_p\left[\sum_{i \in I} f_i(S)\right]\cdot B\\
    &\leq \P_p[S\cap I\neq \emptyset] \cdot B
\end{align*}
with equality for $I = [n]$. Thus, $p$ must satisfy \eqref{eq:condition}.\\

Now, suppose $p\in (0,1)^n$ satisfies \eqref{eq:condition}. For any $f\in \F_{Luce}$ with budget $B$, let $$Z(f):=\max_{i \in [n]} \left(c_i'(b_i(f, p_{-i}))-c_i'(p_i)\right)$$
and 
$$\mathcal{C}(f) := \{ i \in [n] \, : \, c_i'(b_i(f,p_{-i})) - c_i'(p_i) = Z(f) \}.$$

From Equation \eqref{eq:foc_fgn},
\begin{align*}
    \sum_{i \in [n]} p_i \cdot \left[c_i'(b_i(f,p_{-i}))-c_i'(p_i)\right]=0.
\end{align*}

Moreover, for any agent $i\in [n]$,
% \begin{align*}
%     \sum_{j: j \succ i} p_j \cdot c_j^{\prime}\left(b_j\left(f, p_{-j}\right)\right) = B\cdot \E[\sum_{j: j \succ i} f_j(S)] = B \cdot \P[S \cap \{j : j \succ i\} \neq \emptyset] \geq \sum_{j : j \succ i} p_j \cdot c_j'(p_j),
% \end{align*}
% so
\begin{align*}
    \sum_{j: i \succcurlyeq j} p_j \cdot \left[c_j^{\prime}\left(b_j\left(f, p_{-j}\right)\right)-c_j^{\prime}\left(p_j\right)\right] \leq 0.
\end{align*}

It follows that $Z(f)\geq 0$ for all $f\in \mathcal{F}_{Luce}$ and so $$z = \inf_{f \in \mathcal{F}_{Luce}}{Z(f)}\geq 0.$$

We will now show that $z=0$.

Suppose towards a contradiction that $z>0$. Let $f\in \mathcal{F}_{Luce}$ be such that
\begin{enumerate}
\item $Z(f)=z$ and 
\item for any other $g\in \mathcal{F}_{Luce}$ such that $Z(g) = z$, $\mathcal{C}(g)\not \subset \mathcal{C}(f)$. 
\end{enumerate}

Let $(X_1, \dots, X_{\ell})$ be the ordered partition corresponding to $\priority$ and $\lambda_1, \dots, \lambda_n$ be the weights that define $f$.

Let $k$ be the maximum index such that $X_k \cap \mathcal{C}(f) \neq \emptyset$. 

First suppose $k = \ell$. From above, we know that there must be some $i\in X_l$ such that $c_i'(b_i(f, p_{-i}))-c_i'(p_i)<0$. Now consider another Luce contract $g$ which is the same as $f$, except that the weight for agent $i$ is $\lambda_i'=\lambda_i+\epsilon$. Thus, when all agents in $[n]\setminus X_l$ have failed, agent $i$ gets a slightly higher share of the reward if it succeeds under $g$ than it did under $f$.  Notice that for $j \notin X_{\ell}$,
\begin{align*}
    c_j'(b_j(g, p_{-j})) - c_j'(p_j) = c_j'(b_j(f, p_{-j})) - c_j'(p_j)
\end{align*}
and for $j \in X_{\ell} \setminus \{i\}$,
\begin{align*}
    c_j'(b_j(g, p_{-j})) - c_j'(p_j) < c_j'(b_j(f,p_{-j})) - c_j'(p_j) \leq z.
\end{align*}
Moreover, for $\eps$ sufficiently small,
\begin{align*}
    c_i'(b_i(g,p_{-i})) - c_i'(p_i) < 0
\end{align*}
by continuity. It follows that $Z(g) \leq z$. In particular, we either have $Z(g)<z$ or $Z(g)=z$ and $\mathcal{C}(g) \subset \mathcal{C}(f)$. In either case, we have a contradiction.

Thus, it must be that $k < \ell$. Now, consider another Luce contract $g$ which has the partition $(X_1, \dots, X_{k-1}, X_k \cup X_{k+1}, X_{k+2}, \dots, X_{\ell})$ and weights 

$$\lambda_i'= \begin{cases}\lambda_i & \text { if } i \notin X_{k+1} \\ \eps \lambda_i & \text { if } i \in X_{k+1}\end{cases}.$$

Then for $j \notin X_k \cup X_{k+1}$,
\begin{align*}
    c_j'(b_j(g,p_{-j})) - c_j'(p_j) = c_j'(b_j(f,p_{-j})) - c_j'(p_j)
\end{align*}
and for $j \in X_k$,
\begin{align*}
    c_j'(b_j(g,p_{-j})) - c_j'(p_j) < c_j'(b_j(f,p_{-j})) - c_j'(p_j) \leq z.
\end{align*}
Finally, for $\eps$ sufficiently small and $j\in X_{k+1}$,
\begin{align*}
    c_j'(b_j(g,p_{-j})) - c_j'(p_j) < z
\end{align*}
by continuity. It follows that $Z(g) \leq z$. In particular, we either have $Z(g)<z$ or $Z(g)=z$ and $\mathcal{C}(g) \subset \mathcal{C}(f)$. In either case, we have a contradiction.    
Thus, we have that 
$\inf_{f \in \mathcal{F}_{Luce}}{Z(f)}=0$. By compactness of $\mathcal{F}_{Luce}$, there exists an $f\in \mathcal{F}_{Luce}$ such that $Z(f)=0$ which means
$$\left[\max_{i \in [n]} \left(c_i'(b_i(f, p_{-i}))-c_i'(p_i)\right)\right] = 0.$$
But from above, we also know that $$\sum_{i \in [n]} p_i\left[c_i'(b_i(f,p_{-i}))-c_i'(p_i)\right]=0.$$

Therefore, it must be that for all $i\in [n]$, 
$$\left(c_i'(b_i(f, p_{-i}))-c_i'(p_i)\right)=0$$
which implies $f\in E^{-1}(p)$. Thus, for any $p$ that satisfies \eqref{eq:condition}, there is a Luce contract $f$ with budget $B$ such that $p\in E(f)$.

\end{proof}

When possible, Luce contracts provide a particularly desirable alternative for implementation over other contracts by mitigating the uncertainty in total payments. For a contract $f$ with budget $B$ and a profile $p$, denote by $T_{f, B} = \sum_i f_i(S) \cdot B$ the (random) total payment.\\

\begin{thmrep}
\label{prop:mean_spread}
Suppose $p\in (0,1)^n$ is implemented by a Luce contract $f$ with budget $B$ and an FGN contract $g$ with budget $B'$. Then $T_{g, B'}$ is a mean preserving spread of $T_{f, B}$ at $p$.
\end{thmrep}

\begin{proof}
By Proposition~\ref{prop:FGN}, for any FGN contract $h$ (with budget $B$) that implements $p$, $\E_p[T_{h, B}] = \sum_i p_i \cdot c_i'(p_i)$. Moreover, since $h$ is FGN, $S = \emptyset \Rightarrow T_{h, B} = 0$ and $\E_p[T_{h, B}] = \P_p[S \neq \emptyset] \cdot \E_p[T_{h,B} \given S \neq \emptyset]$, so $\E_p[T_{f, B} \given S \neq \emptyset] = \E_p[T_{g, B'} \given S \neq \emptyset]$.\\

Observe that $S \neq \emptyset \Rightarrow T_{f, B} = B$, and it follows that $\E_p[T_{g, B'} \given S \neq \emptyset] = B$. Now, let $X = 0$ if $S = \emptyset$ and $X = T_{g, B'} - B$ otherwise. Then $T_{g, B'} = T_{f, B} + X$, $\E_p[X \given T_{f, B} = 0] = \E_p[X \given S = \emptyset] = 0$, and 
\begin{align*}
    \E_p[X \given T_{f, B} = B] = \E_p[T_{g,B'} - B \given S \neq \emptyset] = 0,
\end{align*}
so $T_{g, B'}$ is a mean-preserving spread of $T_{f,B}$ at $p$.
\end{proof}

The variance of the total payment is minimized by Luce contracts (when they exist). Notice that Luce contracts also minimize the maximum total payment the principal might have to make. In the following example, we illustrate how the reduction in this maximum payment can be significant compared to piece-rate contracts, which are the natural choice of contract in a setting with independent agents and no budget constraint.

\begin{example}
Consider $n$ agents with identical quadratic costs, $c_i(p_i)=\frac{1}{2}Cp^2$, and consider the profile where all agents have the same success probability, $p_i = q$. This profile is implemented by the piece-rate contract which gives every agent $Cq$ if they succeed and nothing otherwise, and it is also implemented by the Luce contract which splits a budget of $B = \dfrac{nCq^2}{1-(1-q)^n}$ equally among all successful agents. 

While the expected payment is $nCq^2$ under both contracts, the maximum possible total payment the principal might have to make is $nCq$ under the piece-rate contract and $\dfrac{nCq^2}{1-(1-q)^n}$ under the Luce contract. In particular, if the principal wants there to be one successful agent in expectation, as might be the case for crowdsourcing an innovation, it would implement $q=\frac{1}{n}$, and the total payment could be as large as $C$ under the piece-rate contract, where as the maximum possible total payment approaches $0$ for large $n$ under the Luce contract.

% Formally, the contract $g_i(S) = \mathbbm{1}_{i \in S} \cdot \frac1n$ with budget $B'=nCq$ corresponds to this piece-rate contract and implements $q$. 

% Formally, the contract $f_i(S) = \mathbbm{1}_{i \in S} \cdot \frac{1}{|S|} \cdot$ with budget $B$ corresponds to this Luce contract and implements $q$.

% The 

% Under $g$, the variance of the payment to any particular agent is $C^2 q^3 (1-q)$, so the variance of the total payment is $n C^2 q^3 (1-q)$. 
% The maximum total payment is $nCq$.

% Under $f$, the variance of the payment is $ n^2C^2q^4 \frac{(1-q)^n}{1-(1-q)^n}$ and 
% the maximum total payment is $\dfrac{nCq^2}{1-(1-q)^n}$.

% If we plug in $q = 1/n$, this becomes

% Under $g$, the variance of the payment to any particular agent is $C^2 q^3 (1-q)$, so the variance of the total payment is $C^2 (1-1/n) /n^2$. The maximum total payment is $C$.

% Under $f$, the variance of the payment is $ n^2C^2q^4 \frac{(1-q)^n}{1-(1-q)^n} \approx (C^2 / n^2) \frac{1}{e-1}$ and the maximum total payment is $\approx \dfrac{C/n}{1-e^{-1}}$.

\end{example}

\section{Conclusion}

We study a contract design problem with multiple independent agents. With binary outcomes for each agent and a budget-constrained principal, we introduce a novel class of contracts, called Luce contracts, and demonstrate their sufficiency for implementing the maximal set of implementable equilibria. It follows that for any objective of the principal, there is always a Luce contract that is optimal. Additionally, we highlight the relevance of Luce contracts in the absence of budget, as they provide a means for inducing effort while reducing the uncertainty in total payments and in particular, mitigating the risk of excessively large payouts.\\

Our results and analysis open several promising directions for future research. Within the framework of our model with binary outcomes, an intriguing question is how the structure of the optimal Luce contract varies with heterogeneity in agents' costs under different objectives of the principal. This is particularly compelling in light of the robustness to cost parameters observed in the special case of two agents with quadratic costs. Another valuable avenue would be to extend our model to allow for multiple outcomes per agent and to investigate potential generalizations of Luce contracts in these richer and more complex environments.

% \nocite{*}
\newpage
\bibliographystyle{ecta}

\bibliography{refs}

\end{document}